\documentclass[aps,prl,twocolumn,floatfix]{revtex4}
\usepackage[pdftex]{color}
\usepackage{latexsym,hyperref,amssymb,amsmath,gensymb}
\usepackage{graphicx,kec}
\usepackage{makeidx}

\newcommand{\om}{\omega}
\newcommand{\by}{\times}
\newcommand{\mb}{\mbox}

\newcommand{\btab}{\begin{tabular}}
\newcommand{\etab}{\end{tabular}}
\newcommand{\barr}{\begin{array}}
\newcommand{\earr}{\end{array}}
\newcommand{\bpm}{\begin{pmatrix}}
\newcommand{\epm}{\end{pmatrix}}
\newcommand{\bit}{\begin{itemize}}
\newcommand{\eit}{\end{itemize}}
\newcommand{\ben}{\begin{enumerate}}
\newcommand{\een}{\end{enumerate}}
\newcommand{\bct}{\begin{center}}
\newcommand{\ect}{\end{center}}

\newcommand{\ra}{\rangle}
\newcommand{\la}{\langle}

\newcommand{\lt}{\left}
\newcommand{\rt}{\right}

\definecolor{navyblue}{rgb}{.05,0,.55}

\begin{document}
\title{Is Dark Energy a Cosmic Casimir Effect?}
\date{\today}
\author{Kevin Cahill}
\affiliation{Department of Physics \& Astronomy\\
University of New Mexico, Albuquerque, New Mexico}
\email{cahill@unm.edu}
\begin{abstract}
Unknown short-distance effects
cancel the quartic divergence
of the zero-point energies.
If this renormalization
took effect in the early universe
after the last phase transition
and applied only to 
modes whose
wavelengths \(\lambdabar\) were shorter than the
Hubble length \(H^{-1}(t^*)\) 
at that time,
then the zero-point energies of
the modes of longer wavelengths
can approximately account for 
the present value of the dark-energy density.
The model makes two predictions.
\end{abstract}
\maketitle
\par
Observations of some 400 type-Ia
supernovas suggest that
the expansion of the universe is 
accelerating~\cite{Riess1998,Perlmutter1999,Frieman2008}
as if subject to a negative pressure.
Negative pressure is the derivative
of the energy with respect to the volume
at constant entropy 
\beq
\mb{} - p = \lt. \frac{\p U}{\p V} \rt|_{S}
\label {p}
\eeq
and so the simplest explanation of this
cosmic acceleration is that the energy density
of empty space is positive.
Data from WMAP and BAO tell us that
this dark-energy density
\(\rho_d\) is about 73 percent 
of the critical density
\(3 H_0^2/8\pi G_N\)
that makes the universe
spatially flat~\cite{Komatsu2011}
and so
\beq
\rho_d = 3.1 \by 10^{-47} \, \mb{GeV}^4.
\label {Ed}
\eeq
\par
Dark energy may be the energy
of the ground state of whatever
fundamental theory describes
the universe.
In quantum field theory
to lowest order in the coupling constants,
the ground-state energy is 
a sum over every type \(\a\)
of elementary field 
and all momenta \(k\)  
of the zero-point 
energies~\cite{Casimir1948,Schwinger1978,Milonni1992,Jaffe2005,Bordag2009}
\beq
E_0 = \sum_{k,\a} (-1)^{2s_\a} \, 
g_\a \, \half \, \hbar \om_{k,\a}
\label {E1}
\eeq
in which \(s_\a\) is the spin
of the particle of field \(\a\), 
\(m_\a\) its mass, 
\(\om_{k,\a} = c \, \sqrt{k^2 + c^2 m_\a^2}\)
its energy, and the statistical weight
\(g_\a\) is \(2s_\a + 1\) if \(m_\a > 0\),
2 if \(m_a = 0\) and \(s_\a > 0\),
and 1 if \(m_a = s_\a = 0\)\@.
\par
The zero-point energy \(E_{0 \a}\)
of every field is badly divergent.
In natural units (\(\hbar = c = 1\)),
it is
\beq
E_{0 \a} =  (-1)^{2s_\a} \, 
g_\a \, \frac{V}{2(2\pi)^3} \!
\int \! \sqrt{k^2 + m_\a^2} \, d^3k.
\label {H00}
\eeq
Inserting a cutoff \(\Lambda\),
we find for a massless scalar field
\beq
E_{0 s} = \frac{V}{16 \pi^2} \, \Lambda^4.
\label {H01}
\eeq
If the cutoff \(\Lambda\)
is the Planck mass
\(m_P = 1.22 \by 10^{19}\) GeV,
then the contribution of a single
massless scalar boson to the
dark-energy density is
\beq
\rho_{s} = 1.4 \by 10^{74} \, \mb{GeV}^4.
\label {rho 0}
\eeq
The ratio of this estimate
to the observed
dark-energy density \(\rho_d\) is
\(
\rho_s/\rho_d = 5 \by 10^{120}
\label {ratio}
\),
making \(\rho_s\) 
too big by more than 120 orders of magnitude.
Clearly, zero-point energies
must either cancel or be
renormalized.
\par
The dark-energy mechanism 
of this paper is based upon two 
related assumptions about 
zero-point energies.
The first assumption is that
at any time \(t\),
an expanding universe 
is sensitive only to the zero-point energies
of wavelengths,
let us say \(\lambdabar = 1/k\),
shorter than the Hubble distance 
\(H^{-1}(t)\) at that time.
The second assumption is that 
at some time \(t^*\)
after the last phase transition
in the early universe, unknown
short-distance effects permanently
renormalized and canceled the 
zero-point energy (\ref{H00})
of the modes to which the
universe was then sensitive,
that is,
modes with wavelengths \(\lambdabar\)
shorter than the Hubble distance
\(H^{-1}(t^*)\) at that time.
These two assumptions imply that
the dark-energy density at later
times involves
only those momenta that lie
within the interval
\(H(t) < k < H(t^*)\) 
\beq
\rho_{C}(t) = \sum_{\a} \frac{(-1)^{2s_\a} \,
g_\a}{16 \pi^3}
\int_{H(t)}^{H(t^*)} \!\! 
\sqrt{k^2 + m_\a^2} \, d^3k.
\label {rhoC}
\eeq
Because the limits on the
momentum integration are both
in the infrared,
the energy \(\sqrt{k^2 + m_\a^2}\)
is dominated by the mass term
except for neutrinos and massless particles.
Thus, the dark-energy density is
approximately
\bea
\rho_{C}(t) & \approx & 
\sum_{\a} \frac{(-1)^{2s_\a} \,
g_\a \, m_\a}{4 \pi^2}
\int_{H(t)}^{H(t^*)} \!\! 
k^2 \, dk \nn\\
& = & \sum_{\a} \frac{(-1)^{2s_\a} \,
g_\a \, m_\a}{12 \pi^2}
\lt( H^3(t^*) - H^3(t) \rt). 
\label {rhoCm}
\eea
\par
If we knew the spectrum of masses
and spins of the elementary
particles and fields,
then we could compute
the Casimir energy density
\(\rho_{C}(t)\) as a function of the
time \(t\)\@.
Instead, I will define
an effective excess \(\la M \ra\)
of boson over fermion
masses as 
\beq
\la M \ra \equiv \sum_{\a} (-1)^{2s_\a} \,
g_\a \, m_\a .
\label {NM}
\eeq
The cosmic Casimir energy density
at time \(t\) is then
\beq
\rho_{C}(t) \approx \frac{\la M \ra}{12 \pi^2}
\lt( H^3(t^*) - H^3(t) \rt). 
\label {rhoCNM}
\eeq
\par
The present value \(H_0\) of the
Hubble constant is so small that
\(\rho_{C}(t_0)\) is effectively
\beq
\rho_{C}(t_0) \approx \frac{\la M \ra \, H^3(t^*)}
{12 \pi^2}. 
\label {rhoCNM0}
\eeq
\par
If the renormalization of the
zero-point energies took place
shortly after the QCD phase transition
at \(t^* = 10^{-5}\) s, then
\(1/H(t^*) = 2 t^* = 2 \by 10^{-5}\) s,
and so
\beq
\rho_{C}(t_0) \approx 
\frac{\la M \ra \, c^2 \, \hbar^3}
{96 \pi^2 \, t^{*3}} = 
3.0 \by 10^{-61} \la M \ra \, c^2
\mb{GeV}^3.
\label {rhoCNM0is}
\eeq
Thus, if the effective excess bosonic
mass (\ref{NM}) were
\beq
\la M \ra \approx 10^{14} \, \mb{GeV}/c^2
\label {NM=}
\eeq
then the present value of
the Casimir energy density would 
approximately equal
the dark-energy density
\beq
\rho_{C}(t_0) \approx \rho_d = 3.1 \by 10^{-47}
\, \mb{GeV}^4.
\label {C=d}
\eeq
\begin{figure}
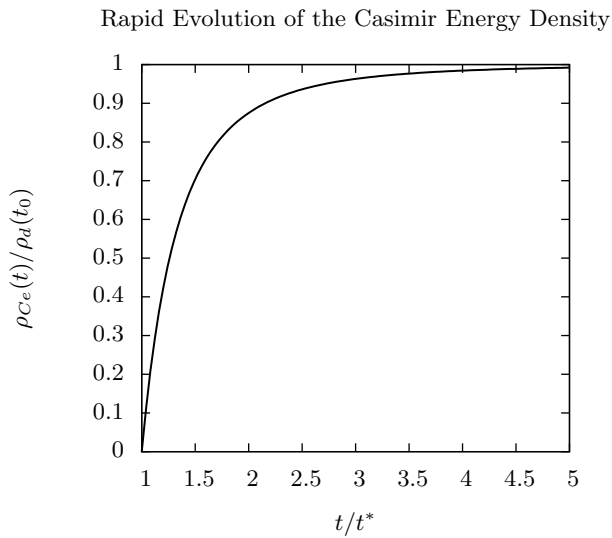

\centering
\input de
\caption{The ratio of the energy
density \(\rho_{C}(t)\) 
of the cosmic Casimir effect (\ref{rhoCNMevoles})
to the dark-energy density \(\rho_d\) 
is plotted
for \(\la M \ra \, c^2 = 1.03 \by 10^{14}\) GeV
in the interval \(1 < t/t^* < 5\)\@.}
\label{defig}
\end{figure}
\par
This explanation of dark energy
makes two predictions.
The first is that the effective
excess \(\la M \ra\) of boson over fermion
masses (\ref{NM}) is \(10^{14}\) GeV/\(c^2\)
which is a plausible order of magnitude
in a theory of grand unification.
\par
The second prediction is that the dark-energy
density varies with time
as in (\ref{rhoCNM}),
rapidly rising from zero at \(t^* = 10^{-5}\) s
to its present value \(\rho_d\),
as in the figure.
It is a kind of quintessence
that does not increase the helium 
abundance~\cite{Weinberg2008.90}\@.
More precisely, 
the very early universe after inflation
is flat and dominated by radiation,
and so the scale factor evolves as
\(a(t) \propto \sqrt{t}\), and the
Hubble parameter as
\(H(t) = \dot a(t)/a(t) = 1/2t\)\@.
Thus, the Casimir energy density (\ref{rhoCNM})
rises as
\beq
\rho_{C}(t) \approx \frac{\la M \ra \, c^2 \hbar^3}
{96 \pi^2}
\lt( t^{*-3} - t^{-3} \rt). 
\label {rhoCNMevoles}
\eeq
\par
If the interval of integration 
is scaled to \(a \, H(t) < k < a \, H(t^*)\)
and the moment of
renormalization to 
\( b \by 10^{-5} \) s,
then the predicted bosonic mass excess
shifts to \( b^3 \la M \ra /a^3 \)\@.

\begin{acknowledgments}
I am particularly grateful
to Peter Milonni for explaining
the Casimir effect
and to Franco Giuliani 
and Randolph Reeder 
for several corrections.
I should also like to thank
R.~Allahverdi, S.~Atlas, B.~Becker,
M.~Hoeferkamp,
and E.~Mottola
for helpful conversations
and M.~Bordag, R.~Jaffe, K.~Milton,
U.~Mohideen, and A.~Parsegian 
for useful e-mail.
\end{acknowledgments}
\bibliography{physics,astro}
\end{document}